\documentclass[
  journal=pasa,
    manuscript=research-article, 
  year=2024,
  volume=37,
]{cup-journal}
\input{preamble}
\usepackage{lineno}
\usepackage{ulem}
\usepackage[final]{changes}
\usepackage{orcidlink}

\usepackage{amsmath}
\usepackage[nopatch]{microtype}
\usepackage{booktabs}
\usepackage{hyperref} 
\hypersetup{colorlinks,citecolor=blue,linkcolor=blue,urlcolor=blue}

\usepackage{orcidlink}

\usepackage[utf8]{inputenc}
\usepackage[nopatch]{microtype}
\usepackage{booktabs}
\usepackage{amsmath} 
\usepackage{caption} 
\usepackage{amsfonts}
\usepackage{amssymb} 
\usepackage{graphicx}
\usepackage{hyperref}
\usepackage{pstool} 
\usepackage{bm} 
\usepackage{multirow}
\usepackage{rotating} 
\usepackage{epstopdf}

\def\vis{\mathcal{V}\left( \mathbf{U}, \nu \right)}

\def\v2{V_2\left(\mathbf{U},\Delta \nu = 0\right)}

\def\dBdTnu{Q_{\nu}}

\def\V{\mathcal{V}}
\newcommand{\vcg}{\mathcal{V}_{cg}}

\newcommand{\HI}{{\rm H\hspace{0.5mm}}{\scriptsize {\rm I}}} 
\newcommand{\cl}{C_{\ell}}
\newcommand{\dnu}{\Delta\nu}

\def\n{\hat{\mathbf{n}}} 
\def\m{\hat{\mathbf{p}}}
\def\c{\hat{\mathbf{c}}}
\def\dn{\Delta \mathbf{n}}
\def\pkp{P(k_{\perp}, k_{\parallel})}
\def\dpc{\Delta {\rm PC}}

\newcommand{\uv}[1]{\hat{\mathbf{#1}}}

\newcommand{\bk}{\mathbf{k}}
 
\newcommand{\kpar}[1][]{k_{\parallel{#1}}}
\newcommand{\alp}{\alpha_p}

\newcommand{\tht}{\theta}

\newcommand{\thf}{\theta_{\rm FWHM}}

\newcommand{\vtht}{\bm{\theta}}
\newcommand{\vch}{\bm{\chi}_p}
\newcommand{\vthtp}{\bm{\theta}-\bm{\chi}_p}
\newcommand{\U}{\mathbf{U}}
\newcommand{\dv}{\mathbf{d}}
\newcommand{\alm}[1][]{a_{\ell}^{m {#1}}}

\newcommand{\kpp}{k_{\perp}}
\newcommand{\eg}{{\it e.g.}\,}
\def\ie{{\it i.e.}\,}

\title[Tracking Tapered Gridded Estimator: I]{The Tracking Tapered Gridded Estimator for the 21-cm power spectrum from MWA drift scan observations I: Validation  and preliminary results} 

\author{Suman Chatterjee\,\orcidlink{0000-0001-8852-5888}}
\affiliation{Department of Physics and Astronomy, University of the Western Cape,
7535 Bellvill, Cape Town, South Africa}
\alsoaffiliation{National Centre for Radio Astrophysics, Tata Institute of Fundamental Research, Post Bag 3, Ganeshkhind, Pune - 411 007, India.}

\email[Suman Chatterjee]{6306414@myuwc.ac.za, sumanchttrj05@gmail.com}

\author{Khandakar Md Asif Elahi\orcidlink{0000-0003-1206-8689}}
\affiliation{Department of Physics, Indian Institute of Technology Kharagpur, Kharagpur - 721 302, India.}

\author{Somnath Bharadwaj\,\orcidlink{0000-0002-2350-3669}}
\affiliation{Department of Physics, Indian Institute of Technology Kharagpur, Kharagpur - 721 302, India.}
\email[Somnath Bharadwaj]{somnath@phy.iitkgp.ac.in}

\author{Shouvik Sarkar\orcidlink{0009-0003-3096-7028}}
\affiliation{Centre for Strings, Gravitation and Cosmology, Department of Physics, Indian Institute of Technology Madras, Chennai 600036, India }

\author{Samir Choudhuri}
\affiliation{Centre for Strings, Gravitation and Cosmology, Department of Physics, Indian Institute of Technology Madras, Chennai 600036, India }

\author{Shiv Sethi}
\affiliation{Raman Research Institute, C. V. Raman Avenue, Sadashivanagar, Bengaluru 560080, India.}

\author{Akash Kumar Patwa\,\orcidlink{0000-0002-6216-2430}}
\affiliation{Raman Research Institute, C. V. Raman Avenue, Sadashivanagar, Bengaluru 560080, India.}


\begin{document}

\begin{abstract}

Drift scan observations provide the broad sky coverage and instrumental stability needed to measure the Epoch of Reionization (EoR)  21-cm signal.  In such observations, the telescope's pointing center (PC) moves continuously on the sky. The Tracking Tapered Gridded Estimator (TTGE)  combines observations from different PC to estimate $P(k_{\perp}, k_{\parallel})$ the 21-cm power spectrum,  centered on a tracking center (TC) which remains fixed on the sky. The tapering further restricts the sky response to a small angular region around TC, thereby mitigating wide-field foregrounds.  Here we consider $154.2 \, {\rm MHz}$ ($z = 8.2$) Murchison Widefield Array (MWA) drift scan observations. The periodic pattern of flagged channels, present in MWA data,  is known to introduce artefacts which pose a challenge for estimating $P(k_{\perp}, k_{\parallel})$. Here we have validated the TTGE using simulated MWA drift scan observations which incorporate the flagged channels same as the data. We demonstrate that the TTGE is able to recover $P(\kpp, \kpar)$ without any artefacts, and estimate    $P(k)$   within $5 \%$  accuracy over a large $k$-range.  We also present preliminary results for a single PC,  combining  9 nights of observation $(17 \, {\rm min}$ total). We find that $P(k_{\perp}, k_{\parallel})$  exhibits streaks at a fixed interval of $\kpar=0.29 \, {\rm Mpc}^{-1}$, which matches $\Delta \nu_{\rm per}=1.28 \, {\rm MHz}$ that is the period of the flagged channels. Since the simulations demonstrate that the TTGE  is impervious to the flagged channels, the streaks seen for the actual data are possibly caused by some systematic that has the same period as the flagged channels. These streaks  are more than  $3-4$ orders of magnitude smaller than the peak foreground power $\mid P(\kpp, \kpar) \mid \approx 10^{16} \, {\rm mK^2}\, {\rm Mpc^3}$    at $\kpar=0$. The streaks are not as pronounced at larger  $k_{\parallel}$, and in some cases they do not appear to extend across the entire $\kpp$ range.  The rectangular region $0.05 \leq \kpp \leq 0.16 \, {\rm Mpc^{-1}}$ and $0.9 \leq \kpar \leq 4.6 \, {\rm Mpc^{-1}}$ is found to be relatively free of foreground contamination and artefacts,  and we have used this to place the $2\sigma$ upper limit $\Delta^2(k) < (1.85\times10^4)^2\, {\rm mK^2}$ on the EoR 21-cm mean squared brightness temperature fluctuations at $k=1 \,{\rm Mpc}^{-1}$. 
 
\end{abstract}

\section{Introduction}

The epoch of reionization (EoR) when the diffuse neutral hydrogen (\HI) in the inter-galactic medium (IGM) underwent a transition to the ionized state is one of the least understood phases in the evolution of our  Universe.
The redshifted  21-cm radiation from \HI{} is a promising observational probe of EoR  \citep{Madau1997, Bharadwaj2005, McQuinn2006, Morales2010, Pritchard2012}.
Several radio interferometers, such as the  Murchison Widefield Array (MWA; 
\citealt{Tingay2013}), LOw Frequency ARray (LOFAR; \citealt{vanHarlem2013}), 
Hydrogen Epoch of Reionization Array (HERA; \citealt{Deboer2017}),  Giant Metrewave Radio Telescope (GMRT; \citealt{Swarup1991, Gupta2017}) and the upcoming SKA-low \citep{Mellema2013, Koopmans2015} aim to detect the power spectrum (PS) of the EoR 21-cm brightness temperature fluctuations.  
Despite substantial observational efforts, it has not been possible to detect the EoR 21-cm PS to date, and we only have upper limits  \citep{Paciga2013, Kolopanis2019, Mertens2020, Trott2020, Pal2020, Abdurashidova2022, Kolopanis2023}. At present we have the best upper limit of  $\Delta^2(k) < (30.76)^{2} \, {\rm mK}^2$  at $k = 0.192\, h\, {\rm Mpc}^{-1}$  for  $z = 7.9$ from HERA \citep{Abdurashidova2022}. 

The main challenge is that the faint \HI{} 21-cm signal is buried in foregrounds which are observed to be three to four orders of magnitude brighter \citep{Ali2008, Bernardi2009, Ghosh2012, Paciga2013, Patil2017}. 
In recent years, there have been significant developments to mitigate the foregrounds, relying upon the fact that foregrounds are spectrally smooth compared to the 21-cm signal. Two approaches are generally taken to deal with the foregrounds. In the `foreground removal' approach, one tries to entirely remove the foregrounds (e.g. \citealt{Chapman2012, Mertens2018, Elahi2023b}). An alternative approach, `foreground avoidance', only uses the $(k_\perp, \kpar)$ region outside the `foreground wedge'  \citep{Datta2010, Morales2012, Vedantham2012, Trott2012, Pober2016} to estimate  $P(k)$ the spherical PS of the EoR 21-cm signal (e.g. \citealt{Dillon2014, Dillon2015b, Trott2020, Pal2020, Abdurashidova2022}). 

The Tapered Gridded Estimator (TGE; \citealt{Choudhuri2014, Choudhuri2016b}) is a visibility-based 21-cm PS estimator that has been extensively used for measuring the 21-cm PS \citep{Pal2020, Pal2022, Elahi2023, Elahi2023b, Elahi2024}, and for characterizing the diffused Galactic foregrounds \cite{Choudhuri2017a} and magnetohydrodynamic turbulence \citep{Saha2019, Saha2021}.  The main attribute of TGE is that it suppresses the sidelobe responses of the telescope to mitigate the effects of extra-galactic point source foregrounds \citep{Ghosh2011a, Ghosh2011b}. Additionally, the TGE is computationally efficient as it deals with gridded visibilities. It is also an unbiased estimator as it internally estimates the noise bias from the self-correlation of the visibilities to yield an unbiased estimate of the PS. In a recent work 
\citet{Chatterjee2022} (hereafter, \citetalias{Chatterjee2022}) have introduced the   Tracking Tapered Gridded Estimator (TTGE), which generalises the TGE for estimating the 21-cm PS from drift scan observations. \citetalias{Chatterjee2022} has also validated the TTGE considering simulated MWA drift scan observations at a single frequency,  where it was demonstrated that the estimated $C_{\ell}$ matches the input model  $C^M_{\ell}$.

Missing frequency channels, fagged to remove  Radio Frequency Interferences (RFI) or for other reasons, pose a serious problem for visibility-based  PS estimation.  The Fourier transform from frequency to delay space (e.g. \citealt{Morales2004, Parsons2009}) introduces artefacts in the estimated PS, and there has been substantial work to address this problem  \citep{Parsons2009, Parsons2014, Trott2016, Kern2021, Ewall-Wice2021, Kennedy2023}. We, however, note that this problem does not arise if we correlate the visibilities directly in the frequency domain \citep{Bharadwaj2001a, Bharadwaj2005} and use this to estimate the 21-cm signal. 
The TGE and the TTGE first correlate the visibilities to estimate $\cl(\dnu)$   the multi-frequency angular power spectrum (MAPS; \citealt{Datta2007, Mondal2018}),  and then Fourier transforms $\cl(\dnu)$ along the $\dnu$ to estimate $P(k_\perp, \kpar)$ the cylindrical PS.  Using simulations,  \cite{Bharadwaj2018} have shown that  $\cl(\dnu)$ does not exhibit any missing $\dnu$ values even when $80 \%$ of randomly chosen frequency channels are flagged in the visibility data,
and it is possible to estimate $P(k_\perp, \kpar)$ without any artefacts due to the missing channels. This has been further borne out in the analysis of actual data using the TGE 
\citep{Pal2020, Pal2022, Elahi2023, Elahi2023b, Elahi2024}. 

MWA has a periodic pattern of flagged channels in the visibility data, which introduce horizontal streaks in  $P(k_\perp, \kpar)$  \citep{Paul2016, Li2019, Trott2020, Patwa2021}. In this paper, we consider the MWA drift scan observation used in \citet{Patwa2021}. Here we investigate if the TTGE can overcome this issue.  For this, we first simulated the MWA drift scan observation with exactly the same flagging as in the actual data and used this to verify if the TTGE can faithfully recover the input model power spectrum used for the simulations.  We have subsequently applied the TTGE to actual observations and present preliminary results here.   

We have arranged the paper in the following manner. First, we describe the MWA drift scan observation in Section~\ref{sec:data}. In Section~\ref{sec:TTGE}, we present the TTGE and the formalism for the power spectrum. Section~\ref{sec:val} describes the simulations and the validation of the TTGE. We have shown the preliminary results from actual MWA observation in Section~\ref{sec:results_data}. We summarize and discuss our findings in Section~\ref{sec:Discussion}.

\section{MWA Drift-scan Observation}\label{sec:data}

\begin{figure}
    \centering
    \includegraphics[width=\columnwidth]{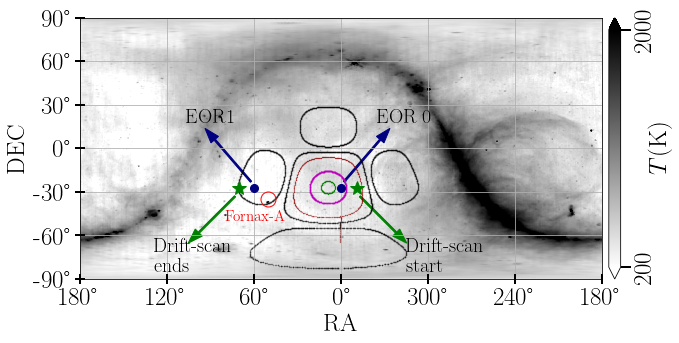}
    \caption{This shows  the  $408 \, {\rm MHz}$  Haslam map \citep{Haslam1982} scaled to 154~MHz assuming the brightness temperature spectral index $\alpha = -2.52$ \citep{Rogers2008}. The iso-contours in green, magenta, red and black show the MWA primary beam at values 0.9, 0.5, 0.05, and 0.005, respectively, for a pointing center at ($6.1^{\circ}, -26.7^{\circ}$) which corresponds to the data analysed here. The scan starts roughly at the location of the `$\star$' on the right (RA=$349^{\circ}$) and lasts until the `$\star$' on the left (RA=$70.3^{\circ}$). Blue filled circles mark the  fields EoR~0($0^{\circ}, -26.7^{\circ}$) and EoR~1($60^{\circ}, -26.7^{\circ}$). The red circle shows the position of Fornax~A.}
    \label{fig:sky_covered}
\end{figure}

The data analysed here is from Phase II (compact configuration) of the Murchison Widefield Array (MWA, \citealt{Lonsdale2009}, \citealt{Wayth2018}) which is a radio interferometer with  128 tiles (antennas in rest of the paper)  located in  Western Australia (latitude $-26.7^\circ$, longitude $116.7^\circ$). Each antenna consists of 16 crossed dipoles placed in a 4x4 arrangement on a square mesh of side  $\sim4~{\rm m}$. MWA operates in several frequency bands ranging from 80~MHz to 300~MHz.  The observed visibilities are recorded  with the time resolution of 0.5~s, and they are written out at  an  interval of 2~minutes (one snapshot) in which the actual duration of observation is  112~s. 

For this work we consider a particular drift scan observation (project ID G0031) that is described in \citet{Patwa2021}. In short, the observation is at a fixed declination (DEC)  $-26.7^{\circ}$ which corresponds to the zenith and it covers a  region of the sky from right ascension (RA) $349^{\circ}$ to $70.3^{\circ}$ which spans $81.3^{\circ}$ in RA over a time duration of  5~hr 24~min. The beginning and end of the drift scan observation are marked with $\star$ in Figure~\ref{fig:sky_covered}. Visibilities are recorded every 2~min which leads to $162$ different pointing centers (PCs, labelled  PC=1,2,...,162),  located at an interval of $0.5^{\circ}$ along RA. 
This region includes two well observed MWA fields namely  EoR~0 ($0^{\circ}, -26.7^{\circ}$) and EoR~1 ($60^{\circ}, -26.7^{\circ}$) which are also shown in Figure~\ref{fig:sky_covered}. The same drift scan observation was carried out on  10 consecutive nights.  The present observation has been performed at a nominal frequency of $\nu_c=154.2 \, {\rm MHz}$  (nominal redshift $z_c=8.2$) with $N_c = 768$ channels of resolution of $\dnu_c=40 \, {\rm kHz}$  covering the observing bandwidth of $B_{\rm bw} = 30.72$ MHz. This is further divided into  24 coarse bands each containing  32 channels or $1.28 \, {\rm MHz}$.

\begin{figure}
    \centering
    \includegraphics[width=\columnwidth]{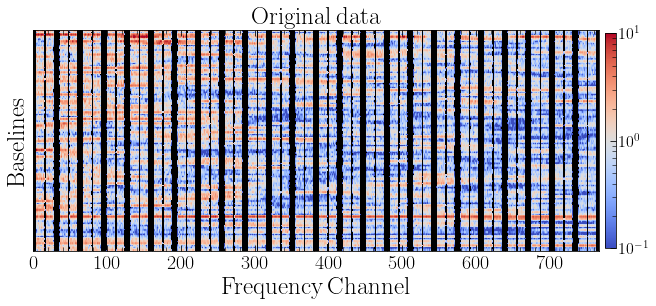}
    \caption{This shows the periodic channel flagging in the observed MWA visibility data. The entire frequency bandwidth is divided into $24$ coarse bands of $32$ channels or $1.28 \, {\rm MHz}$ width each. 
    The colors here shows arbitrarily normalised visibility amplitudes, and the  black vertical lines indicate the flagged channels.}
    \label{fig:bl-nu}
\end{figure}

The data has been pre-processed with COTTER (\citealt{Offringa2015}) which flags RFI and  non-working antennas. For each coarse band, COTTER also flags   four channels 
at both ends and one channel at the  center resulting in channels (1-4,17,29-32) to be flagged. The produces a period pattern of flagged channels as shown in Figure~\ref{fig:bl-nu}. We apply COTTER individually to the 0.5 s time resolution visibility data, and then average them to 10~s time resolution. This pre-processed data is written in \texttt{CASA}\footnote{\url{https://casa.nrao.edu/}} readable \textit{Measurement Sets} (MS). The  MS for 10 nights are calibrated separately using the steps mentioned in \citet{Patwa2021}. 
We note that the first $2 \, {\rm hr}$ of data is missing from the $6$-th night, and as a consequence the nights of observations is  $N_{\rm nights}=10$  for some PCs  whereas it is $N_{\rm nights}=9$ for others.  Since the observations covers the same region of the sky, we perform Local Sidereal Time (LST) stacking (\eg \citealt{Bandura2014, Amiri2022}) and obtain the equivalent  one night  drift scan data. 
We finally have $162$ MS,  each corresponding to a different pointing direction on the sky. Each MS contains visibility data with $11$ different time stamps each with $t_{\rm int}= N_{\rm nights} \times 10  \, {\rm s}$ effective integration time. 

In addition to the sky signal, each measured visibility also has a system noise contribution. For the present data, $\sigma_{\rm N}$ the r.m.s. system noise  for the real (and also imaginary) part of the  measured visibilities  is estimated using  
\begin{eqnarray}
    \sigma_{\rm N} = \frac{1}{\eta_s} \frac{2 K_{\rm B} T_{\rm sys} }{A_{\rm eff} \sqrt{\dnu t_{\rm int}}}
    \label{eq:nrms}
\end{eqnarray}
where the $A_{\rm eff} = 21.4 \, {\rm m}^2$ and $T_{\rm sys} / \eta_s = 290 {\rm K}$ \citep{Patwa2021}. This yields   $\sigma_{\rm N} = [N_{\rm nights}]^{-0.5} \times 60 \, {\rm Jy}$ for a single visibility. 

Considering the baseline distribution, $99$ percent of the baselines are found to be smaller than 500~m, i.e., $257.8\lambda$ (Figure 2 of \citetalias{Chatterjee2022}),  and we have restricted the baseline range in our analysis to $\mid U \mid \leq 245 \lambda$.

\section{The Tracking TGE}\label{sec:TTGE}

$T(\uv{n}, \nu)$ the brightness temperature distribution on the sky is decomposed into spherical harmonics  $Y_{\ell}^{m}(\uv{n})$  using
\begin{equation}
    T(\uv{n}, \nu) = \sum_{\ell, m}\alm(\nu) Y_{\ell}^{m}(\uv{n})
    \label{eq:sh}
\end{equation}
where  $\alm(\nu)$ are the expansion coefficients. The MAPS which is defined as  
\begin{equation}
     \cl(\nu_a,\nu_b) = \langle\alm(\nu_a) \alm[*](\nu_b) \rangle \, ,
     \label{eq:cl3}
\end{equation}
jointly characterizes the angular and frequency dependence of the two-point statistics of $T(\uv{n}, \nu)$. Further, the MAPS is a function of the frequency separation 
\begin{equation}
    \cl(\nu_a, \nu_b) = \cl(\dnu) \,\,  {\rm where} \,\,   \Delta \nu=  \mid \nu_b-\nu_a\mid 
    \label{eq:erg}
\end{equation}
if $T(\uv{n}, \nu)$ is assumed to be statistically homogeneous  (ergodic) along the line of sight direction.

\citetalias{Chatterjee2022} presents the tracking tapered gridded estimator (TTGE) to determine the MAPS $\cl(\nu_a,\nu_b)$ using the measured visibility data from drift scan radio-interferometric observations.
\citetalias{Chatterjee2022} also explained how  $\cl(\nu_a,\nu_b)$ can be used to  determine the cylindrical power spectrum $P(k_{\perp},k_{\parallel})$.  
However, the validation in that paper is restricted to a single frequency $\nu$. 
There, we have simulated observations where the sky signal $T(\n)$ corresponds to an input model $C^M_{\ell}$. We have applied TTGE on the simulated visibilities to estimate $C_{\ell}$ and demonstrated that the estimated $C_{\ell}$ matches the input model.
In this section of the present paper, we briefly summarize the formalism of the TTGE, and in Section~\ref{sec:val}, we validate it considering multi-frequency observations and the full three-dimensional power spectrum $P(k)$. 

The visibility $\vis$ measured at a baseline $\U$ and frequency $\nu$ is given by,  
\begin{equation}
\vis = \dBdTnu \int d\Omega_{\n}
T\left(\n,\nu \right) A\left(\Delta \n,\nu
\right) e^{2 \pi i \U\cdot \dn },
\label{eq:v1}
\end{equation}
where $\dBdTnu = 2 k_B / \lambda^2$ is the conversion factor from  brightness temperature to specific intensity in the Raleigh-Jeans limit,  $T\left(\hat{\mathbf{n}},\nu \right)$ is the brightness temperature distribution on the sky,  $d\Omega_{\n}$ is the elemental solid angle in the direction $\n$, $A\left(\Delta \hat{\mathbf{n}},\nu \right)$ is the antenna primary beam (PB) pattern, and $\Delta \n =\n -\m$  where  $\m$ is the telescope's pointing direction or pointing center (PC).  In this work, we consider drift scan observations where the telescope is held fixed (on earth) to point towards the zenith. Here   $\m$  has a fixed declination $\delta_0$ (latitude of the array), whereas the right-ascension $\alp$ varies due to the earth's rotation. In such a situation, it is convenient to use   $\mathcal{V}(\alp,\U_i,\nu_a)$ where we have included  $\alp$ as a parameter which tells us the pointing direction $\m$ for which the visibilities were recorded. As noted earlier, for the observations considered here, we have $\delta_0=-26.7^{\circ}$ and $349^{\circ} \le \alp \le 70.3^{\circ}$ at an interval $\Delta \alp =0.5^{\circ}$.

The question is ``How do we combine the visibility data from different $\m$?''.  To address this, we choose a tracking center (TC) $\c$ which remains fixed on the sky and  consider the convolved visibilities defined in the $uv$  plane as 
\begin{equation}
    \mathcal{V}_c(\alp,\U,\nu)=\int d^2 U^{'} \, \tilde{w}(\U-\U^{'}) \, e^{2 \pi i \U^{'} \cdot \vch } \,  \mathcal{V}(\alp,\U^{'},\nu) \, 
    \label{eq:f8}
\end{equation}
where $\vch=\m -\c$  and  $\tilde{w}(\U)$  is the Fourier transform of a  tapering function   $W\left(\vtht \right)$  which is typically chosen to be considerably narrower than the antenna's primary beam pattern $A\left(\vtht,\nu \right)$. Here we have used  a Gaussian $ W\left(\vtht \right)=e^{-\theta^2/\theta_w^2}$.
  Adopting the flat-sky approximation, eq.~(\ref{eq:f8})  can be expressed as 
\begin{equation}
    \mathcal{V}_c(\alp,\U,\nu)=\dBdTnu \int d^2 \tht \, 
T\left(\vtht,\nu \right) \, W(\vtht) \, A\left(\vthtp,\nu
\right) e^{2 \pi i \U \cdot \vtht } 
\label{eq:f9}
\end{equation}
where the phase center of $\mathcal{V}_c(\U,\nu)$ is shifted to  $\c$, and the function $W(\vtht)$ restricts the sky response to a small region around  $\c$ .  For the purpose of this discussion, we may assume that the convolved visibility $\mathcal{V}_c(\alp,\U,\nu)$   only records the signal from a small region of the sky centered around $\c$.  We can now coherently combine the data from different pointings 
\begin{equation}
    \mathcal{V}_c(\U,\nu)= \sum_p s_p \mathcal{V}_c(\alp,\U,\nu)
\label{eq:f10}
\end{equation}
to estimate the sky signal from a small region around  $\c$. The contribution to $\mathcal{V}_c(\U,\nu)$ from  $\mathcal{V}_c(\alp,\U,\nu)$
at a particular PC $\m$ goes down as  $\sim  A\left(-\vch,\nu\right)$ (eq.~\ref{eq:f9})  which  declines  rapidly as the separation $\vch=\m-\c$ increases. In addition to the sky signal, each $\mathcal{V}_c(\alp,\U,\nu)$  also contains a system noise contribution.   This implies that the PC  $\m$, which are close to $\c$ contribute to  $\mathcal{V}_c(\U,\nu)$    with a relatively high signal-to-noise ratio (SNR) as compared to those which are at a large angular distance from $\c$.  We account for this by suitably choosing the factor $s_p$, which assigns different weights to every  PC. 

Considering the observational data, we evaluate the convolved visibilities on a rectangular grid (labelled using  $g$) on a $uv$ plane. Note that our analysis does not account for the fact that the baseline $\U = \dv \nu/c$  corresponding to a fixed antenna separation $\dv$ changes with frequency, and it is held fixed at the value corresponding to $\nu_c$.
The convolved gridded visibilities are evaluated using 
\begin{align}
    \vcg(\nu_a) =  \sum_{p}  s_p \sum_{n} & \tilde{w}(\U_g-\U_i) e^{2 \pi i \U_i \cdot \vch } \times   \nonumber \\
    & \mathcal{V}(\alp,\U_i,\nu_a) F_{p, n}(\nu_a) \,
\label{eq:c1} 
\end{align}
where $F_{p, n}(\nu_a)$ has a value $0$ if the particular visibility is flagged and is $1$ otherwise. Here the sum over different baselines $\U_i$ is evaluated for a fixed PC (labelled $p$), whereas the outer sum combines all the different PCs covered in the drift scan observation. The final 
$\vcg(\nu_a)$ refers to the sky signal centered at a particular TC $\c$. 

Following  eq.~(6) of \cite{Pal2022}, we define the tracking tapered gridded estimator (TTGE) as  
\begin{align}
    \hat{E}_g(\nu_a,\nu_b) = & M^{-1}_g(\nu_a,\nu_b) {\mathcal Re} \,  \Big{[} \vcg(\nu_a)   \mathcal{V}^*_{cg}(\nu_b)   \nonumber \\
    & - \sum_{p,i} F_{p, i}(\nu_a) F_{p, i}(\nu_b) \mid s_p  \tilde{w}(\U_g-\U_i) \mid^2 \nonumber \\
    &  \mathcal{V}(\alp,\U_i,\nu_a) \mathcal{V}^*(\alp,\U_i,\nu_b) \Big{]}
\label{eq:c2} 
\end{align}
where $M_g(\nu_a,\nu_b)$ is a  normalisation factor and ${\mathcal Re}[]$ denotes the real part. The second term in the {\it r.h.s.} of eq.~(\ref{eq:c2})  subtract  out the contribution from the correlation between the visibilities measured at the same baseline and pointing direction.  This is primarily introduced to cancel out the noise bias which arises when we correlate a visibility with itself ({\it i.e.} $a=b$; \citetalias{Chatterjee2022}). 

We use all-sky simulations (Section \ref{sec:sim}) to estimate $M_g(\nu_a,\nu_b)$.  The simulated sky signal   $T(\uv{n},\nu)$ is a realisation of a  Gaussian random field corresponding to an unit multi-frequency angular power spectrum (UMAPS) where  $\cl(\nu_a,\nu_b)=1$. The simulations consider identical drift scan observations as the actual data, with  exactly the same flagging, and frequency and baseline coverage. The simulated visibilities $ [\mathcal{V}(\alp,\U_i,\nu_a)]_{\rm UMAPS}$ are  analysed  exactly the same as the actual data. We have estimated $M_g(\nu_a,\nu_b)$ using 
\begin{align}
    & M_g  (\nu_a, \nu_b)= \Big{\langle } {\mathcal Re} \,  \Big{[} \vcg(\nu_a)   \mathcal{V}^*_{cg}(\nu_b) - \sum_{p,n} F_{p, n}(\nu_a) F_{p, n}(\nu_b) \times \nonumber \\
    & \mid s_p  \tilde{w}(\U_g-\U_i) \mid^2 
      \mathcal{V}(\alp,\U_i,\nu_a) \mathcal{V}^*(\alp,\U_i,\nu_b) \Big{]}    
    \Big{\rangle}_{\rm UMAPS} 
    \label{eq:c3}
\end{align}
where the angular brackets $\langle ... \rangle$ denote an average over multiple realisations of the  UMAPS. Here we have used $100$ random  realisations of the UMAPS  to reduce the statistical uncertainties in the estimated  $M_g(\nu_a,\nu_b)$.

The  MAPS TTGE defined in eq.~(\ref{eq:c2}) 
provides an unbiased estimate of $\cl{_g}(\nu_a,\nu_b)$ at the angular
 multipole $\ell_g=2 \pi U_g$ {\it i.e.} 
\begin{equation}
\langle {\hat E}_g(\nu_a,\nu_b) \rangle = \cl{_g}(\nu_a,\nu_b)
\label{eq:a1}
\end{equation}
To enhance the SNR and also reduce the data volume, we have divided the $uv$ plane into annular bins.  We  use this to define the binned TTGE 
\begin{equation}
\hat{E}_G[q](\nu_a,\nu_b) = \frac{\sum_g w_g  {\hat E}_g(\nu_a,\nu_b)}
{\sum_g w_g } \,.
\label{eq:a2}
\end{equation}
where $w_g$ refers to the weight assigned to the contribution from any particular 
grid point $g$. For the analysis presented in this paper we have used the 
weight $w_g=M_g(\nu_a,\nu_b)$  which roughly averages the visibility correlation 
$\V_{cg}(\nu_a) \,  \V_{cg}^{*}(\nu_b)$ across  all the  grid points
 which are sampled by the baseline distribution and lie within the boundaries of bin $q$.

The binned estimator  has an expectation value 
\begin{equation}
\bar{C}_{\bar{\ell}_q}(\nu_a,\nu_b)
  = \frac{ \sum_g w_g \cl{_g}(\nu_a,\nu_b)}{ \sum_g w_g}
\label{eq:a3}
\end{equation}
where $ \bar{C}_{\bar{\ell}_q}(\nu_a,\nu_b)$ is the bin averaged MAPS 
 at 
 \begin{equation}
\bar{\ell}_q =
\frac{ \sum_g w_g \ell_g}{ \sum_g w_g}
\label{eq:a4}
\end{equation}
which is the effective angular multipole for bin $q$. For brevity of notation, we use
$C_{\ell}(\nu_a,\nu_b)$  instead of $\bar{C}_{\bar{\ell}_q}(\nu_a,\nu_b)$ in the subsequent discussion. 

For the subsequent analysis, we assume that the 21-cm signal is statistically homogeneous  (ergodic) along the line of sight whereby $\cl(\nu_a, \nu_b) = \cl(\dnu)$ (eq.~\ref{eq:erg}). Such an assumption is valid for the redshifted 21-cm signal if the observing bandwidth $B_w$ is sufficiently small 
($\approx 8 \, {\rm MHz}$; \citealt{ Mondal2018}) such  that the mean neutral hydrogen fraction does not evolve significantly across the corresponding redshift interval. Although 
$B_w = 30.72 \, {\rm MHz}$ used here is quite a bit larger and we may expect some signal loss at the small $k$, we still adopt this assumption as it significantly simplifies the analysis as we can quantify the signal using $P(k_{\perp},\,k_{\parallel})$  the cylindrical PS  of the 21-cm brightness temperature fluctuations which is related to a $\cl(\Delta \nu)$ through a Fourier transform. We have \citep{Datta2007}  
\begin{equation}
    P(k_{\perp},\,k_{\parallel})= r^2\,r^{\prime} \int_{-\infty}^{\infty}  d (\Delta \nu) \,
    e^{-i  k_{\parallel} r^{\prime} \Delta  \nu}\, \cl(\Delta \nu)
    \label{eq:cl_pk}
\end{equation}
where $k_{\parallel}$ (the Fourier conjugate of $\Delta \nu$) and  and $k_{\perp}=\ell/r$  are  component of $\bk$ respectively parallel and perpendicular to the line of sight. Here 
 $r = 9210 \,{\rm Mpc}$ is the comoving distance corresponding to $\nu_c$  and $r^{\prime}=d r/d \nu = 16.99 \,{\rm Mpc\, MHz}^{-1}$  evaluated at $\nu_c$. The reader is referred to \citet{Bharadwaj2018} for further details. 
 
To proceed further with the TTGE, we have collapsed the 
measured  $\cl(\nu_a, \nu_b)$ to obtain $C(\Delta \nu_n)$ where $\Delta \nu_n =n \, \Delta \nu_c$ with $n=0,1,2,...,N_c-1$. We then use a maximum likelihood estimator (MLE; \citealt{Pal2022}) to obtain  the cylindrical PS 
\begin{equation}
    P(k_{\perp}, k_{\parallel q})   =  \sum_n \left[ \left(\textbf{A} ^{\dagger} \textbf{N}^{-1} \textbf{A}\right)^{-1} \textbf{A}^{\dagger} \textbf{N}^{-1} \right]_{qn} \mathcal{W}(\Delta\nu_n) \, C_{\ell}(\Delta\nu_n)
    \label{eq:ML}
\end{equation}
where the Hermitian matrix $\textbf{A}$ contains the Fourier coefficients and $\textbf{N}$ is the noise covariance matrix. We have assumed $\textbf{N}$ to be diagonal, and estimated it  using noise only simulations. In these simulations the real and imaginary parts of the measured visibilites  were both replaced with   Gaussian random noise  of variance $\sigma_{\rm N}^2$. These simulated data were run through the TTGE pipeline to estimate $C_{\ell}(\Delta\nu_n)$, and we have used $20$ realisations of the simulated visibilities to estimate $[\delta\cl(\dnu_n)]^2$ the variance of $C_{\ell}(\Delta\nu_n)$ which are the diagonal elements of $\textbf{N}$. Note that TTGE avoids the noise bias, and the mean $C_{\ell}(\Delta\nu_n)$ is expected  to be  zero for noise only simulations. 

 The window function $\mathcal{W}(\Delta\nu_n)$, which is normalized to unity at $\Delta\nu = 0$, is used to avoid a discontinuity in the measured $C_{\ell}(\Delta\nu_n)$ at the band edges. For the present work we have  used  a Blackman-Nuttall (BN; \citealt{Nuttall1981}) window function.

\section{Validating the TTGE}
\label{sec:val}

\subsection{All-sky simulation}
\label{sec:sim}

The aim here is to simulate the visibilities that would be measured in  MWA  drift scan observations of  $T(\n,\nu)$ the redshifted 21-cm  brightness temperature distribution on the sky. We simulate $T(\n,\nu)$ using the package \texttt{HEALPix} \citep[Hierarchical Equal Area isoLatitude Pixelization of a sphere;][]{Gorski2005}. 
For the simulations presented here, we have set $N_{\rm side}=512$ in \texttt{HEALPix}, which results in $\ell_{max}= 1535$ and a pixel size of $6.87^{'}$. 

We assume $T(\n,\nu)$ to be a Gaussian random field (GRF) with a given input model power spectrum $P^m (\kpp, \kpar)$. To simulate the signal, we consider 
\begin{equation}
    \cl(\dnu_n) = \frac{1}{N_c  \dnu_c \, r^2\,r^{\prime}} \sum_{q=0}^{N_c-1}
    P(k_{\perp}, k_{\parallel q}) \, e^{2 \pi i n q /N_c}  \, 
    \label{eq:cl2}
\end{equation}
which is the discrete representation  of the inverse of eq.~(\ref{eq:cl_pk}). We generate  
the expansion coefficients $\alm(\nu_n)$  at the $n$-th frequency channel    (eq.~\ref{eq:cl3}) using 
\begin{equation}
    \alm(\nu_n) = \sum_{q=0}^{N_c-1} \left[ \sqrt{\frac{P^m(\kpp, \kpar[q])}{N_c \dnu_c \, r^2 \, r^{\prime}  }} \left( \frac{\uv{x}_q + i \uv{y}_q}{\sqrt{2}} \right)\right]\, e^{\frac{2 \pi i n q}{N_c}} \,
    \label{eq:alm1}
\end{equation}
where $\uv{x}_q,\uv{y}_q$ are independent Gaussian random variables  of unit variance, and 
use eq.~(\ref{eq:sh}) to calculate  $T(\n,\nu_n)$. We now use eq. (4) of  \citetalias{Chatterjee2022} to calculate the simulated visibilities $\mathcal{V}(\alp,\U_i,\nu_a)$ for all the PCs covered in the actual data. Note that we have simulated the visibilities for the same baseline distribution, frequency channels, and flagging of the actual data. For the present work we have assumed the  model PS to be 
\begin{equation}
    P^m (k) = (k/k_0 )^s \, {\rm K}^2 \, {\rm Mpc}^3
    \label{eq:modps} \, 
\end{equation}
with $k_0 =1 \, {\rm Mpc}^{-1}$ and $s = -1$. We have used $20$ independent relaisations of the simulated signal to estimate the mean and $1 \sigma$ errors for all the results presented below. 

In order to simulate the sky signal corresponding to UMAPS, we first consider a fixed frequency channel (say $n=0$) and generate a GRF $T(\n,\nu_0)$ corresponding to a unit angular power spectrum  $C_{\ell}=1$.  We then assign the same brightness temperature map to all the other frequency channels within our observing bandwidth {\it i.e.} $T(\n,\nu_n)=T(\n,\nu_0)$  for $n = 1,2,3,...,N_c-1$. This ensures that the simulated sky signal $T(\n,\nu_n)$ corresponds to $\cl(\nu_a,\nu_b)=1$ (UMAPS). The UMAPS visibilities  
$ [\mathcal{V}(\alp,\U_i,\nu_a)]_{\rm UMAPS}$  were simulated using \texttt{HEALPix} exactly the same way as for the model PS, with the exception that the sky signal is different. As mentioned earlier, we have used $100$ realisations of UMAPS to estimate $M_g(\nu_a,\nu_b)$.
 
We have introduced noise in the simulations as a GRF with zero mean and standard deviation $\sigma_{\rm N} = 10 \, {\rm Jy}$. This is equivalent to an observation where $N_{\rm nights} = 36$ (eq.~\ref{eq:nrms}) and sets the ${\rm SNR}=3$ at the smallest $k$-bin for the input model considered here. The actual EoR signal is significantly fainter and requires much longer observations for a detection.


\subsection{Single pointing}
\label{sec:single_PC}
 
\begin{figure}
    \hspace*{-0.25cm}
    \includegraphics[scale = 0.45]{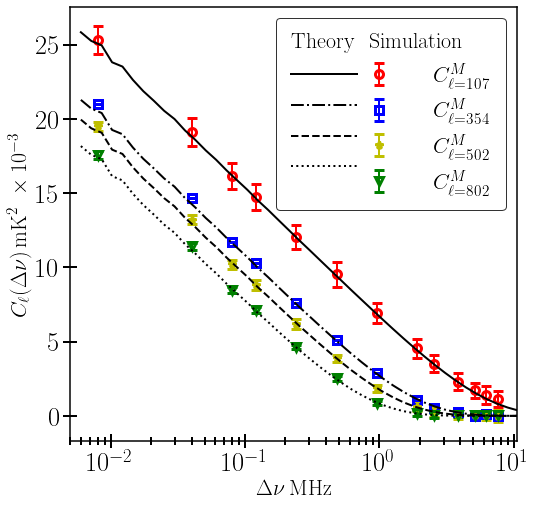}
        \caption{This shows $\cl(\dnu)$ as a function of $\dnu$ for four values of $\ell$. The data points with $1\sigma$ error bars are estimated from 20 realizations of the all-sky simulations. The lines show the theoretical predictions calculated using the input model power spectrum $P^m(k)=(1 {\rm Mpc}^{-1}/k) \, {\rm K^2}\, {\rm Mpc^3}$ in equation~(\ref{eq:cl_pk}). The $\dnu = 0$ points have been slightly shifted for the convenience of plotting on a logarithmic scale. } 
    \label{fig:valid_cl}
\end{figure}

In this subsection, and also the subsequent subsection, we use the simulated visibilities  to validate the TTGE considering a single tracking center 
(TC) at $({\rm RA,DEC)} = (6.1^{\circ},-26.7^{\circ})$. 
We have used a window function  (eq.~\ref{eq:f9}) $ W\left(\vtht \right)=e^{-\theta^2/\theta_w^2}$ where $\theta_w =0.6 \, \theta_{\rm FWHM}$ with $\thf = 15^{\circ}$ . This effectively restricts the sky signal to an angular region of extent $\sim 15^{\circ}$ centered around the TC, and this remians fixed even as the pointing center PC drifts across the sky.  In this section, we have considered a single PC (=34) which exactly coincides with the TC.  Note that the simulations used here do not contain any system noise. 

Figure~\ref{fig:valid_cl} shows the estimated MAPS $\cl(\dnu)$ along with the model predictions $\cl^{\rm M}(\dnu)$. Note that the  range 
$ 50 < \ell < 1493  $ has been divided into  20 $\ell$ bins, and the results are shown for $4$ representative  bins. We see that model predictions are within the $1 \sigma$ error bars of the  estimated $\cl(\dnu)$, indicating that the two are  in good agreement. It is important to note that  the estimated  $\cl(\dnu)$  shows a smooth $\Delta \nu$ dependence with no missing frequency separations  $\Delta \nu$.  In particular, we do not see any artefacts in the estimated $\cl(\dnu)$ due to the periodic pattern of flagged  channels  present in the simulations (Figure \ref{fig:bl-nu}).

\begin{figure}
    \hspace*{-0.5cm}
    \includegraphics[scale = 0.35]{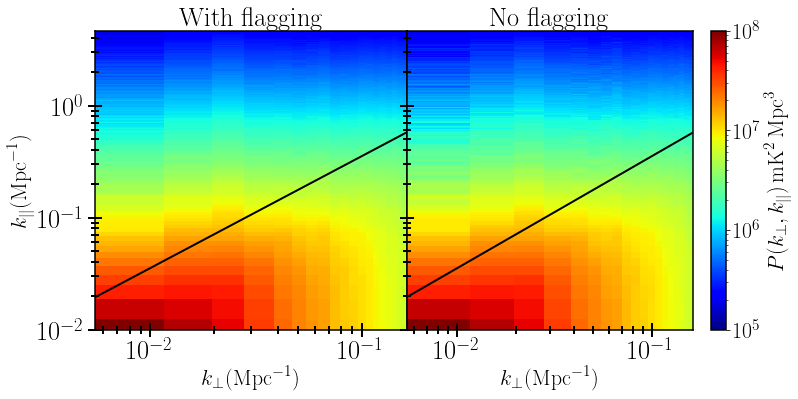}
    \caption{Left panel shows the cylindrical power spectrum  $P(\kpp, \kpar)$ estimated from simulations with MWA coarse channel flagging. For comparison, the right panel shows the $P(\kpp, \kpar)$ estimated from simulations without coarse channel flagging. We do not notice any significant difference.}
    \label{fig:valid_3pk}
\end{figure}

The left panel of Figure~\ref{fig:valid_3pk} shows the cylindrical PS  $P(\kpp, \kpar)$ estimated from  the $\cl(\dnu)$ shown in Figure~\ref{fig:valid_cl} using eq.~(\ref{eq:ML}). For comparison, the right panel shows $P(\kpp, \kpar)$ estimated using identical simulations which do not incorporate the channel flagging.  We note  that the $P(\kpp, \kpar)$ shown in the two  panels  are visually indistinguishable. 
Several  earlier works which have first transformed from  frequency to delay space and then estimated the PS (e.g. \citealt{Paul2016, Li2019, Trott2020, Patwa2021})  have reported horizontal streaks in the estimated $P(\kpp, \kpar)$ due to the periodic pattern of flagged  channels present in the 
MWA data.  In our approach \citep{Bharadwaj2018}, we first estimate $\cl(\dnu)$, which does not have any  missing $\dnu$ even when the visibility data  contains flagged channels. We then Fourier transform  $\cl(\dnu)$ to obtain  a clean estimate of  $P(\kpp, \kpar)$. We see that the missing frequency channels do not introduce any artefacts in the estimated $P(\kpp, \kpar)$.

\begin{figure}
\hspace{-0.25cm}
    \includegraphics[scale = 0.35]{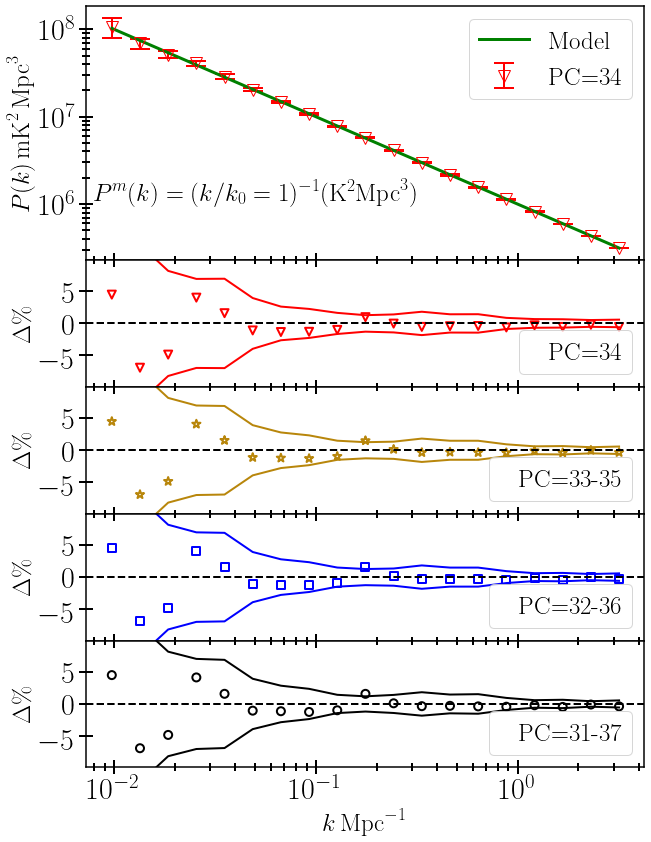}
    \caption{ The upper panel shows the estimated spherically binned power spectrum $P(k)$ and $1-\sigma$ error bars for simulations for PC=34 with no noise and coarse channel flagging. For comparison, the input model $P^m(k)=(1 {\rm Mpc}^{-1}/k) \, {\rm K^2}\, {\rm Mpc^3}$  is also shown by the solid line. The lower panels show the percentage error $\Delta= [P(k) - P^m (k)]/P^m (k)$ (data points) and the relative statistical fluctuation $\sigma /P^m (k) \times 100 \%$ (between the solid lines). The four lower panels consider situations for combining different PCs mentioned in the figure legends.}
    \label{fig:valid_1pk}
\end{figure}

The top panel of figure~\ref{fig:valid_1pk} shows $P(k)$ the spherical  PS estimated directly from the $\cl(\dnu)$  shown in Figure~\ref{fig:valid_cl} using a maximum likelihood estimator (MLE)  \citep{Elahi2023}, for comparison we also show $P^m(k)$ the input model PS.   We see that  the model predictions $P^m(k)$ are within the $1 \sigma$ error bars of the estimated  $P(k)$, indicating that the two are in good agreement through the entire $k$ range 0.01 to 4 ${\rm Mpc}^{-1}$ probed here.  We have quantified the percentage  deviation  between $P(k)$ and $P^m(k)$ using $\Delta = [P(k) - P^m(k)]/P^m(k) \times 100$ shown in the second panel (from top) of figure~\ref{fig:valid_1pk}. 
We see that for nearly all  $k$ the values of $\Delta$ are within the predicted $1 \sigma$ error bars. The values of $\mid \Delta \mid$ are within $1 \%$ for $k \ge  0.13 \, {\rm  Mpc^{-1}}$ 
and within $2 \%$ for $k \ge   0.035 \, {\rm  Mpc^{-1}}$. We see that  
 $\mid \Delta \mid$ increases at smaller $k$ due to the convolution with the window function and the primary beam pattern. We have the largest deviation   $|\Delta| = 6.9 \%$ at $  k = 0.013 \,  {\rm Mpc^{-1}} $.  Overall, we find very good agreement between   $P(k)$ and $P^m(k)$. Further,  we do not find any artefacts due to the periodic pattern of flagged channels present in the MWA data.

\begin{figure}
\centering
\includegraphics[scale = 0.35]{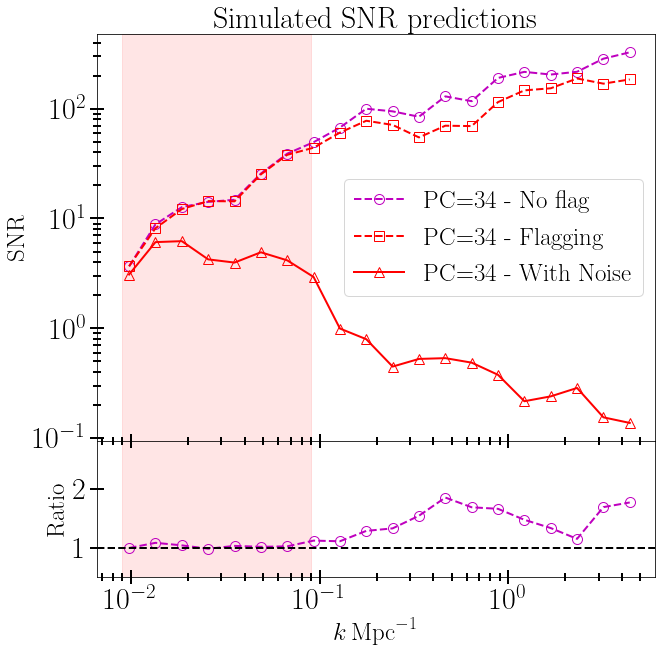}
    \caption{The upper panel shows a comparison of SNR achievable for a single pointing with (circles) and without (squares) the periodic pattern of flagged channels.  The triangles show the expected SNR values in the presence of system noise with $\sigma_{\rm N} = 10 \, {\rm Jy}$ (eq.~\ref{eq:nrms}). The lower panel shows the ratio of the SNR values without and with flagging. The shaded region indicates the $k$-range where the SNR values remain mostly unaffected due to flagging.
    }
    \label{fig:FNF}
\end{figure}

The flagged channels present in the MWA data cause $\sim 28 \%$  data loss, and we expect this to degrade the SNR of the  21-cm PS relative to the situation when there are no flagged channels. The upper panel of Figure~\ref{fig:FNF} shows a comparison of the  SNR for the 21-cm PS estimated from the simulations with (squares) and without (circles) the periodic pattern of flagged channels. In the absence of flagging, the SNR has value $\sim 4$ at the smallest $k$ bin and rises monotonically to $\sim 300$ at the largest $k$ bin. The results with flagging are similar, but the SNR values are somewhat smaller. 
The lower panel shows the ratio of the SNR without flagging to with flagging.   We find that the ratio is very close to unity at $k < 0.09 \, {\rm Mpc}^{-1}$, and the ratio increases only at large $k$ where it varies in the range $1$ to $2$. We find that The ratio has a maximum value $(\sim 2)$ at $k\sim 0.5 \, {\rm Mpc}^{-1}$. Note that the discussion, till now, has not considered the system noise.   The upper panel of  Figure~\ref{fig:FNF}  also shows the SNR for the simulations which include the system noise. We see that the SNR at the two smallest $k$ bins are not much affected even if we introduce the system noise. The total noise budget in these two bins is dominated by the cosmic variance {\bf (CV)}, and the system noise make only a small contribution. We find that there is a very substantial drop in the SNR at the larger $k$ bins when we introduce the system noise. There is very little difference in the SNR between with and  without flagging once we introduce the system noise contribution.

\subsection{Multiple pointings}
\label{sec:multi}

\begin{figure}
    \centering
    \includegraphics[scale = 0.35]{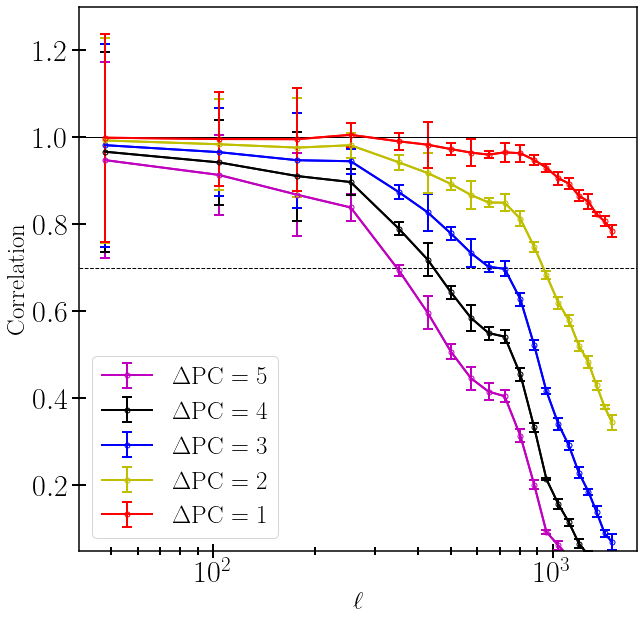}
    \caption{This shows a measure of visibility correlation between two different PCs for a fixed TC. The correlation = $\cl(\dnu=0, \Delta {\rm PC})/\cl(\dnu=0, \Delta {\rm PC}=0)$. Here we have fixed the TC at PC=34 and estimated the MAPS $\cl(\dnu)$ by correlating visibilities separated by $\Delta {\rm PC}$ pointing centers. Here we only show the results for $\dnu = 0$, however, $\dnu > 0$ results are quite similar.}
    \label{fig:cf}
\end{figure}

In this subsection we coherently combine the visibilities measured at multiple PCs to estimate the signal in a small angular region centered 
at the fixed TC $\c$ whose sky coordinates we have mentioned in the previous subsection.  As mentioned earlier, the contribution from a PC  declines as  $\sim  A\left(-\vch,\nu\right)$ (eq.~\ref{eq:f9}) as the separation $\vch=\m-\c$ increases.  The PC which are close to the TC contribute to $\vcg(\nu_a)$ with the higher SNR as compared {\bf to} the PC which are at a large angular separation, and we account for this by suitably choosing the factor $s_p$ (eq.~\ref{eq:c1}).

In the previous subsection we have considered a single pointing center PC=34 which exactly coincides with the TC {\it i.e.} $\m=\c$ and $\vch=0$. In this subsection we consider  other pointing directions PC=34 + $\Delta$ PC.  Before combining the signal from multiple pointing directions, we analyse how the correlation between the signal from two different pointing directions PC=34 and PC=34 + $\Delta$ PC changes  as $\Delta \, {\rm PC} $ is varied.  To evaluate this we consider the dimensionless correlation coefficient 
\begin{equation}
    c_{\ell}(\Delta \, {\rm PC}) = [C_{\ell}(0)]_{\Delta \, {\rm PC}} /C_{\ell}(0)
    \label{eq:corr}
\end{equation}
where we have evaluated $[C_{\ell}(0)]_{\Delta \, {\rm PC}}$ using eq~(\ref{eq:c2}) with the difference that $\vcg(\nu_a) $  and $  \mathcal{V}^*_{cg}(\nu_b)$  refer to PC=34 and  
PC=34 + $\Delta$ PC respectively.  Note that these simulations do not contain system noise, and it is not necessary to subtract out the self-correlation in eq~(\ref{eq:c2}). 

Figure~\ref{fig:cf} shows $c_{\ell}(\Delta \, {\rm PC}) $ as a function of $\ell$ for different values of $\Delta \, {\rm PC}$ in the range $1$ to $5$.  We see that in all cases, the correlation drops as $\Delta \, {\rm PC} $, the offset between TC and PC, is increased. This is roughly consistent with the expected  $\sim  A\left(-\vch,\nu\right)$ (eq.~\ref{eq:f9}) decline, and also the finding of \citet{Patwa2019}. This analysis sets the choice for combining different PC later in the analysis.

Considering $\Delta \, {\rm PC}=1$, we see that $c_{\ell}(\Delta \, {\rm PC}) \approx 1$ for $\ell \le 300$, it is $>0.9$ for $\ell \le 800$ and it drops at larger $\ell$ to $\sim 0.8$ at $\ell = 2,000$. We find a similar behaviour for larger $ \Delta \, {\rm PC}$, but the   values of $c_{\ell}(\Delta \, {\rm PC}) $ are smaller. Overall, the signal at small $\ell \,  (\le 300)$ remains correlated for large $\Delta \, {\rm PC}$, even beyond $5$. However, at larger $\ell$ $(> 800)$ the correlation falls below $\sim 0.7$ for  $ \Delta \, {\rm PC} \ge 3$. We have also investigated $c_{\ell}(\Delta \, {\rm PC}) $   for other values of $\Delta \nu$ (eq.~\ref{eq:corr}), and we find that the behaviour is very similar to that for $\Delta \nu=0$ shown here. We note that the variation of $c_{\ell}(\Delta \, {\rm PC}) $ with $\Delta \, {\rm PC}$ depends on the width of the window function where we have used  $\thf = 15^{\circ}$, and we expect the correlation to decrease faster with $ \Delta \, {\rm PC}$ if $\thf$ is reduced.

The results shown in Figure~\ref{fig:cf} indicate that it is most optimal to consider the weights $s_p$ in eq.~(\ref{eq:c1}) as  functions of $\ell$. However, we have not considered this here. We have considered  three simple schemes where we use uniform wights $s_p=1$ to combine multiple pointings. In the first scheme we combine  PC=33, 34 and 35 (33-35) for which the maximum  $ \mid \Delta \, {\rm PC} \mid $ has value 1. We similarly also consider PC= 32-36 and  31-37 for which the maximum $ \mid \Delta \, {\rm PC} \mid$ has values  $2$ and $3$ respectively. The correlation at large $\ell$ falls considerably for  
  $ \mid \Delta \, {\rm PC} \mid \ge 4$, and we have not considered combining such pointings.   Considering PC=31-37, the RA difference between PC=31 and PC=37 is $6^{\circ}$ and the total time duration is $14 \, {\rm min}$.

The $\cl(\Delta \nu)$, $\pkp$ and $P(k)$  estimated after combining multiple pointings are very similar to those for a single pointing, and we have not shown these separately.  Considering the three schemes for combining multiple pointings, the three lower panels of Figure~\ref{fig:valid_1pk} show $\Delta$ the percentage deviation between the estimated $P(k)$ and the input model $P^m(k)$.  We see that the results from the three schemes are practically indistinguishable from those for a single pointing. Note that this is expected as the simulations considered here do not have any system noise. We expect the system noise to drop when multiple pointings are combined, however, we do not expect the sky signal or the cosmic variance to change. 

\begin{figure*}
    \hspace*{-1.03cm}
    \includegraphics[scale = 0.35]{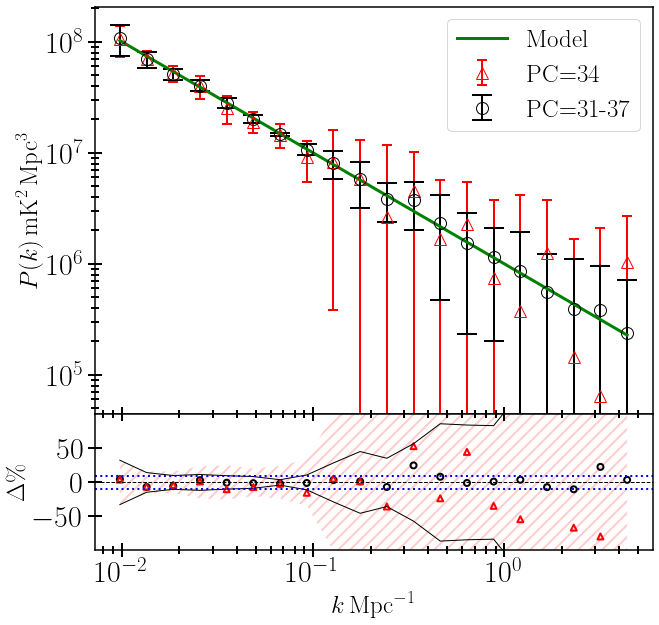}
    \includegraphics[scale = 0.35]{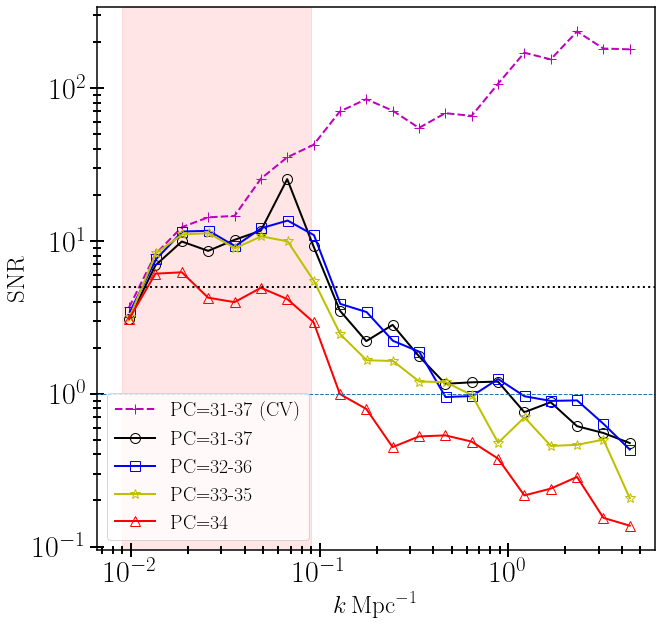}
    \caption{ Left upper panel shows the estimated spherical PS $P(k)$ after combining multiple pointing centers coherently. The solid line shows the input model ($P^m(k)=(1 {\rm Mpc}^{-1}/k) \, {\rm K^2}\, {\rm Mpc^3}$) used for the simulations. 20 realizations of the simulations are used to estimate the mean and $1 \sigma$ errors shown here.  The red triangles and the error-bars show the results for PC=34, whereas the black circles and error-bars correspond to PC=31-37 \ie all the pointing centers between PC=31 and 37 have been combined. Considering the  $P(k)$ estimates in the left upper panel,  the left lower panel shows the corresponding percentage deviation relative to  the input model, the hatched regions (in red) and the region between black solid line show the $1\sigma$  statistical fluctuations for PC=34 and PC=31-37 respectively. Blue dotted lines show $|\Delta|=10\%$. The right panel shows the  SNR after combining multiple PCs.  Here, PC=31-37(CV) refers to the SNR in the absence of system noise, where we only have cosmic variance (CV). This is very close to the SNR for a single pointing (PC=34 - Flagging) shown in the right panel of  Figure~\ref{fig:FNF}.}
    \label{fig:cf_comb_nw}
\end{figure*}

Figure~\ref{fig:cf_comb_nw} shows the estimated $P(k)$ and the SNR when the simulations include system noise. Considering the left panel, we see that the estimated $P(k)$ is consistent with the input model $P^m(k)$ within the $1\sigma$ error bars through  the entire $k$ range.  However, the predicted $1 \sigma$ error bars  and the percentage deviation from $P^m(k)$ depend on the number of PCs that we combine. For a single PC  we find that   $\mid\Delta\mid < 10 \%$ for  $k<0.2 \, {\rm Mpc}^{-1}$ whereas it increases up to  $50\%$ at larger $k$. Considering multiple pointings where we combine PC=31-37, we find that the $1 \sigma$ error bars reduce considerably, and $\mid\Delta\mid<10\%$ for the entire $k$ range. 
The right panel of Figure~\ref{fig:cf_comb_nw} shows how the SNR improves as we combine multiple PCs. Note that this is very similar to the right panel of Figure~\ref{fig:FNF} which shows the SNR for a single pointing PC=34.
For reference,  PC=31-37(CV) shows the SNR in the absence of system noise, where we only have cosmic variance. We do not expect the cosmic variance to change when we combine multiple pointings, and the SNR for PC=31-37(CV) is very close to that for `PC=34-Flagging' shown in the right panel of Figure~\ref{fig:FNF}. Considering the results with system noise,
we note that the first $2$ $k$ bins are cosmic variance (CV) dominated, and these  SNR values do not change when we introduce system noise or combine PCs. The SNR  in all the other $k$ bins degrades substantially when we introduce system noise. We see that we have the lowest  SNR   when we consider  a single PC as compared to multiple PCs. For a single PC, the SNR peaks in the second and third $k$ bins where  ${\rm SNR}= 6$, the SNR falls at larger $k$ and it is below unity at $k>0.09\,{\rm Mpc}^{-1}$.  The SNR improves significantly as we combine multiple  PCs. Combining PC = 33-35, we can achieve ${\rm SNR} \sim 10$ in the range $0.02<k<0.07\,{\rm Mpc}^{-1}$ and the SNR is greater than unity up to $k \sim 0.6\,{\rm Mpc}^{-1}$. Considering PC = 32-36, we find ${\rm SNR}\sim10$ in the range $0.02<k<0.09\,{\rm Mpc}^{-1}$ and the SNR is greater than unity up to $k \sim 1\,{\rm Mpc}^{-1}$. We find that the SNR  improves further for  PC=31-37.  However, this improvement is very small,  which indicates that $\dpc = 3$ is indeed an optimal choice for combining multiple PCs.

\section{Preliminary results from the  observed data} 
\label{sec:results_data}
 
In this section, we present the results from the actual MWA observation described in Section~\ref{sec:data}. The analysis is restricted to a single pointing PC=34 which we have  considered in section~\ref{sec:single_PC}. For this PC, we have $N_{\rm Nights}=9$ whereby  $\sigma_N = 20 \, {\rm Jy}$ (eq. \ref{eq:nrms}). 
Considering the measured visibilities, we find that the amplitudes at small baselines are much larger than those at the longer baselines. The estimated  $\cl(\dnu)$  is expected to be correlated across an extent $\delta \ell \sim 40$ (and possibly larger) due to the tapering  (Figure 5,  \citealt{Chatterjee2022}), and there is a risk of the high foreground level at the small baselines leaking out to the larger baselines.  To avoid this, we have discarded the baselines shorter than $6\lambda$ for the subsequent analysis. We also note an earlier work 
\citep{Li2019},  which has discarded the baselines shorter than $12 \lambda$  to avoid foreground leakage into the EoR window. 
 We have applied the TTGE on the data to estimate $\cl(\dnu)$,  for which the results are presented in ~\ref{app:cl}. We have used these $\cl(\dnu)$ values to estimate the cylindrical power spectrum $P(\kpp, \kpar)$ which we now discuss in some detail. 

\begin{figure*}
    \includegraphics[scale=0.555]{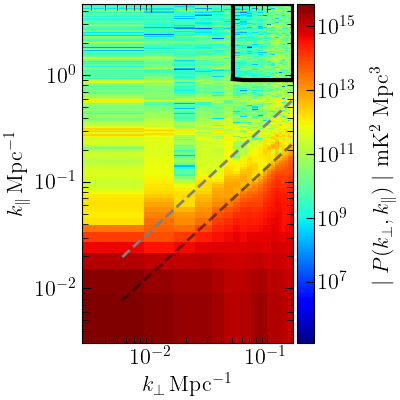}
    \includegraphics[scale=0.555]{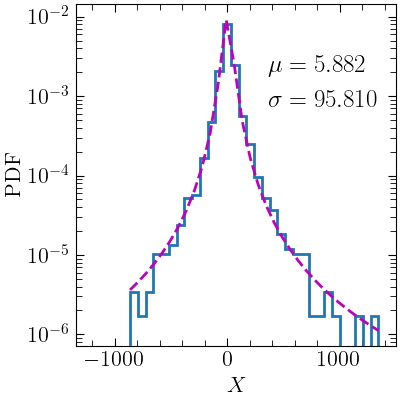}
    \includegraphics[scale=0.555]{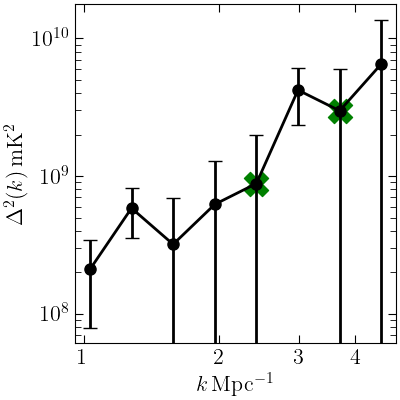}
    \caption{The left panel shows the cylindrical PS $\mid P(\kpp, \kpar) \mid$. The grey and black dashed lines show the theoretically predicted boundary of foreground contamination expected from a monochromatic source located at the horizon and the FWHM of the telescope's PB, respectively. The region inside the black rectangle is used to constrain the 21-cm signal. The middle panel shows the histogram of the quantity $X=P(\kpp, \kpar) / \delta P_{N}(\kpp, \kpar)$ considering the modes inside the rectangle. The right panel shows $\mid \Delta^2(k) \mid$ the absolute values of the mean squared brightness temperature fluctuations and the corresponding $2\sigma$ error bars. The negative values of $\Delta^2(k)$ are marked with a cross.}
    \label{fig:pc4}
\end{figure*}

The left panel of Figure~\ref{fig:pc4} shows the estimated $\mid P(\kpp, \kpar) \mid$ that is  found to have values in the range $10^7 - 10^{15} \, {\rm mK^2}\, {\rm Mpc^3}$. The grey dashed line shows the boundary of the foreground wedge, which corresponds to the foreground contamination expected from a monochromatic source located at the horizon. The black dashed line shows the same for a source located at $\theta = 23^{\circ}$, which corresponds to the FWHM of the telescope's PB. We find that the large values $\mid P(\kpp, \kpar) \mid > 10^{15}  \, {\rm mK^2}\, {\rm Mpc^3}$, which are due to the foregrounds, are localised within the FWHM line.   We also notice a relatively smaller foreground contribution $(\mid P(\kpp, \kpar) \mid \sim 10^{13}  \, {\rm mK^2}\, {\rm Mpc^3})$ between the FWHM line and the wedge boundary. In addition to this, we also find a significant amount of foreground leakage beyond the wedge boundary, and the values of $\mid P(\kpp, \kpar) \mid$ gradually decrease to $\le 10^{11}  \, {\rm mK^2}\, {\rm Mpc^3}$ at $\kpar \sim 0.3\,{\rm Mpc}^{-1}$ and beyond. 

\begin{figure}
    \includegraphics[width=\columnwidth]{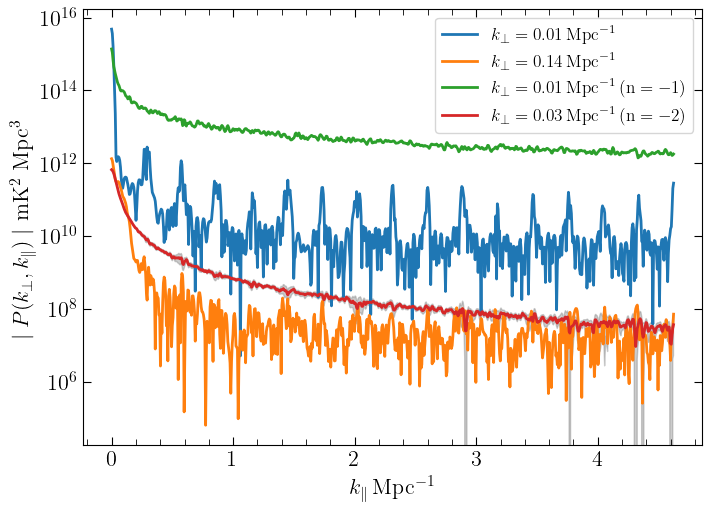}
    \caption{ This figure shows $\mid P(\kpp, \kpar) \mid$ as a function of $\kpar$ for fixed values of $\kpp$. The blue and orange curves are from the observed data. The green and red curves are from simulations with $P(k) \propto k^{-1}$ and $P(k) \propto k^{-2}$, respectively.} The curve corresponding to $\kpp = 0.14\,{\rm Mpc}^{-1}$ {\bf (for the data)} has been divided by a factor of $10^3$, while  {\bf the curves corresponding to the simulations have been scaled arbitrarily for better visualization. The grey shaded region shows $2\sigma$ uncertainties for the simulation with $P(k) \propto k^{-2}$. } 
    \label{fig:per_sys}
\end{figure}

We notice a few horizontal streaks in the power spectrum that extend across several  $\kpp$ bins at some fixed values of $\kpar$. In particular, we find two very closely spaced horizontal streaks at $\kpar \approx 0.29 \, {\rm Mpc^{-1}}$ where $\mid P(\kpp, \kpar) \mid \sim 10^{12} \, {\rm mK^2}\, {\rm Mpc^3}$ that is in excess of the values ($\mid P(\kpp, \kpar) \mid \sim 10^{11} \, {\rm mK^2}\, {\rm Mpc^3}$) in the neighbouring $k_{\parallel}$ bins. Note that this is the lowest $\kpar$ value where we have a horizontal streak. The horizontal streaks at larger  $k_{\parallel}$ are not as pronounced, and in some cases they do not appear to extend across the entire $\kpp$ range. In order to understand these better, we consider Figure~\ref{fig:per_sys} which shows $\mid P(\kpp, \kpar) \mid$ as a function of $\kpar$ for two fixed values of $\kpp$   ($=0.01$ and  $0.14\,{\rm Mpc}^{-1}$).  Considering  $k_\perp=0.01\,{\rm Mpc}^{-1}$, which is the lowest $k_\perp$ bin,  we find spikes in the power spectrum at a regular interval of $\Delta \kpar \sim 0.29\,{\rm Mpc}^{-1}$. The power in these peaks is roughly of the order of $10^{11} - 10^{12} \, {\rm mK^2}\, {\rm Mpc^3}$, which is nearly an order of magnitude larger than the adjacent values of $\mid P(\kpp, \kpar) \mid$. However note that these spikes are nearly $3-4$ orders of magnitude smaller than the peak foreground power $\mid P(\kpp, \kpar) \mid \approx 10^{16} \, {\rm mK^2}\, {\rm Mpc^3}$   that  occurs at $\kpar=0$. We find that  $\kpar  = 0.29 \, {\rm Mpc^{-1}}$,  which  is   the position of the first spike and also the spacing between the successive spikes, corresponds to a period of $\dnu_{\rm per}  = 1.28  \,{\rm MHz}$ in the measured $\cl(\dnu)$. Note that this is exactly the spacing between the coarse bands of MWA  (Figure~\ref{fig:bl-nu}), which is also the period of the flagged channels. We find that for several $\ell$ values, the measured  $\cl(\dnu)$ (Figure~\ref{fig:cl_sys}) exhibits tiny oscillatory features that become clearly visible once we fit and subtract out the dominant smoothly varying component. These oscillatory features, which are shown in the inset of the panels of Figure~\ref{fig:cl_sys},  turn out to have a period of $\dnu_{\rm per}  = 1.28  \,{\rm MHz}$. Note that the amplitude of these oscillatory components is around two orders of magnitude smaller than the total measured $\cl(\dnu)$.  We can attribute the spikes seen in $\mid P(\kpp, \kpar) \mid$ for $k_\perp=0.01\,{\rm Mpc}^{-1}$ ( Figure~\ref{fig:per_sys}) to the oscillatory features in  $\cl(\dnu)$ for $\ell=51$.  We next consider  $k_\perp=0.14\,{\rm Mpc}^{-1}$ which is the $17$-th $k_\perp$ (and $\ell$) bin. In this case, we do not see such prominent spikes as those seen for  $k_\perp=0.01\,{\rm Mpc}^{-1}$.  Considering $\cl(\dnu)$ for $\ell=1273$, here also we see that the oscillatory pattern is not so well defined as for $\ell=51$. For comparison, Figure~\ref{fig:per_sys}   shows $\mid P(\kpp, \kpar) \mid$ for $k_\perp=0.01\,{\rm Mpc}^{-1}$  considering the  simulations  where $P(k) \propto k^{-1}$ ($s=-1$ in eq.~\ref{eq:modps}),  for which the results are shown in Figure \ref{fig:valid_3pk}. The results from  these simulations, which have exactly the same periodic pattern of flagged channels as the actual MWA data, do not show any of the spikes and other artefacts seen in the corresponding results for the actual data. However, the results for these simulations only exhibits a 2 order of magnitude dynamic range, whereas the periodic structures in the data  exhibits a 4 order of magnitude dynamic range with respect to $k_{\parallel}=0$. In order to achieve a higher dynamic range in the simulations, we have also considered $P(k) \propto k^{-2}$ for which the results are also shown in Figure~\ref{fig:per_sys}.
The simulation now  exhibits a 4 order of magnitude dynamic range, for which we still do not see the spikes. We however notice enhanced statistical fluctuations at some of the $k_{\parallel}$ 
where we have spikes in the actual data, which arise because these $k_{\parallel}$ values are poorly sampled due to the periodic pattern of flagged channels.  We further note that the large dynamic range in the actual data is a direct consequence of the fact that foregrounds that dominate these observations are essentially two dimensional (in the plane of the sky), and their 
smooth spectral behaviour leads to a  large value at $k_{\parallel}=0$. It is difficult to replicate this in the three dimensional simulations that we have considered here. We plan to consider foreground simulations in future work. 

The results from our  simulations  suggests that the spikes and other artefacts seen in the measured $\mid P(\kpp, \kpar) \mid$ presented here are not just a straightforward consequence of the periodic pattern of flagged channels. Our results seem to indicate that the spikes are possibly caused by some systematic effect in the data, which has the same period as the missing channels, periodic systematic errors in the calibration being one such possibility. 
We further note that it may be possible to model and remove these artefacts by subtracting out small period components from the measured $\cl(\dnu)$ using Gaussian process regression or some similar technique \citep{Mertens2018, Elahi2023b}.
 
We next attempt to constrain the $z_c=8.2$  EoR 21-cm signal using the measured $P(\kpp, \kpar)$. 
To avoid significant foreground contamination, it is necessary to restrict the region of the $(\kpp, \kpar)$  plane which is used to estimate the 21-cm signal. Here,  we have limited the 
$k_{\parallel} $ range to choose a region that is well above the foreground wedge. We further wish to avoid the large spikes seen in the measured  $\mid P(\kpp, \kpar) \mid$.  We find that these spikes are relatively more prominent  in the first six $\kpp$ bins where they extend over the entire $\kpar$ range. In contrast, the spikes are not so prominent in the subsequent $\kpp$ bins where they also do not extend out to the large  $\kpar$ range.  This feature is visible when we compare the results for $k_\perp = 0.01$ and  $0.14\,{\rm Mpc}^{-1}$ in  Figure~\ref{fig:per_sys}. 
Based on the above considerations we have discarded the first six   $\kpp$ bins and  chosen the rectangular region $0.05 \leq \kpp \leq 0.16 \, {\rm Mpc^{-1}}$ and $0.9 \leq \kpar \leq 4.6 \, {\rm Mpc^{-1}}$, which is shown in the top right corner of the left panel of Figure~\ref{fig:pc4}. 
The subsequent analysis is restricted to this rectangular region where  
 $\mid P(\kpp, \kpar) \mid $ is found to have values in the range $10^7 - 10^{11} \, {\rm mK^2}\, {\rm Mpc^3}$.
 
We now consider the statistics of the measured $P(\kpp, \kpar)$, particularly to assess whether these values can be utilized to meaningfully constrain the EoR 21-cm signal. Following \citet{Pal2020}, we consider the   quantity 
\begin{equation}
    X=\frac{P(\kpp, \kpar)}{\delta P_{N}(\kpp, \kpar)} \, 
    \label{eq:xstat}
\end{equation}
which is the ratio of the measured $P(\kpp, \kpar)$ to $\delta P_{N}(\kpp, \kpar)$   the statistical uncertainty expected from the system noise contribution only. We expect $X$ to have a symmetric distribution with mean $\mu = 0$ and standard deviation $\sigma_{\rm Est} = 1$ if the estimated values of $P(\kpp, \kpar)$ are consistent with the uncertainties predicted due to the system noise contribution only.  Residual foreground contamination will manifest itself as a positive mean, whereas a negative mean would indicate some sort of negative systematics in the power spectrum estimates. A value $\sigma_{\rm Est} > 1$ would indicate the presence of additional sources of uncertainty, beyond the predicted system noise contribution.

The middle panel of Figure~\ref{fig:pc4} shows the probability density function (PDF) of $X$, which is found to be largely symmetric around $0$ with $\mu=5.88 $ and  $\sigma_{\rm Est}=95.81 $.   Here, the substantially large value of $\sigma_{\rm Est}$   indicates that  the measured 
$P(\kpp, \kpar)$ has fluctuations which are considerably larger than those predicted from the system noise contribution alone. We note that several earlier works have reported such `excess variance'  for the estimated 21-cm PS at $\sim 150 \, {\rm MHz}$ \citep{Mertens2020, Pal2020}  and also at higher frequencies $ \sim 430 \, {\rm MHz} $  \citep{Pal2022,Elahi2023, Elahi2023b,Elahi2024}. The exact cause of this excess variance is not known at present. We note that  the  earlier works which have applied the TGE \citep{Pal2020,Pal2022,Elahi2023,Elahi2023b,Elahi2024} have obtained values of $\sigma_{\rm Est}$ in the range $2$ to $5$, whereas the value  $\sigma_{\rm Est}=95.81 $  obtained here  is more than an order of magnitude  larger.  Although the exact cause is not known,  it is plausible that the systematics  responsible for the spikes in the PS at the lower values  of  $\kpp$ and  $\kpar$ also  causes excess fluctuations in  $P(\kpp, \kpar)$ within the rectangular region used to constrain the 21-cm signal.  
We find that the mean $\mu=5.88 $ is not consistent with zero within the expected statistical fluctuations predicted using $\sigma_{\rm Est}$.  This indicates that the measured $P(\kpp, \kpar)$ has some residual foreground contamination which introduces a positive bias in the distribution if $X$. 
We further note that the PDF of $X$ is not Gaussian, and it nicely fit by  a Lorentzian distribution with  a peak location $x_0 = 1.68$ and spread $\gamma = 2.73$, consistent with the earlier studies  \citep{Elahi2023, Elahi2023b,Elahi2024} all of which find that $X$  follows a Lorenztian distribution.  For the subsequent analysis, we have scaled the system noise only error predictions with $\sigma_{\rm Est}$ to account for the excess variance {\it i.e.} we have used $ \delta P(\kpp, \kpar) =  \sigma_{\rm Est} \times   \delta P_{N}(\kpp, \kpar)  $ to predict the statistical fluctuations expected in the measured $P(\kpp, \kpar)$. 

We use the $(\kpp, \kpar)$ modes in the rectangular region mentioned earlier to estimate the spherical power spectrum $P(k)$. The right panel of Figure~\ref{fig:pc4} shows the mean squared brightness temperature fluctuations $\Delta^2(k) = k^3 P(k)/(2 \pi^2)$ as a function of $k$ along with the $2\sigma$ error bars corresponding to the predicted statistical fluctuations $\delta P(\kpp, \kpar)$.  The expected statistical fluctuations are much larger than the EoR  21-cm signal which is predicted to typically have values of the order of  $\Delta^2(k) \sim 10^1-10^2 \, {\rm mK}^2 $ over the range $k=1 - 4 \, {\rm Mpc}^{-1} $ considered here \citep{Mondal2017}. In the absence of foreground contamination, we expect $\Delta^2(k)$  to fluctuate around $0$ with both positive and negative values that are consistent with the predicted statistical fluctuations. We find that  $\Delta^2(k)$ in most of the $k$-bins are consistent with the statistical fluctuations, and the values are within  $0\pm2\sigma$. However, the values of $\Delta^2(k)$ in the first, second and sixth $k$-bins are somewhat larger, and these are consistent with the statistical fluctuations only at $0\pm5\sigma$. It is possible that these $k$-bin may have some residual foreground contamination, however, it is also possible that these are more extreme statistical fluctuations in the extended tail of the Lorentzian distribution. Note that the two $k$-bins with negative $\Delta^2(k)$ values are both consistent with  $0\pm2\sigma$ indicating that the estimated PS is free from any negative systematics. Since the measured PS is largely consistent with the statistical fluctuations and there are no negative systematics, we use the measured values to place $2\sigma$ upper limits on  $\Delta^2(k)$. The tightest upper limit is found to be $\Delta^2(k) < (1.85\times10^4)^2\, {\rm mK^2}$ at the first $k$-bin $k=1 \,{\rm Mpc}^{-1}$.

\section{Summary and Discussion}
\label{sec:Discussion}

Drift scan observations provide an economic and stable option with the broad sky coverage required for 21-cm EoR experiments. In this paper, we consider the Tracking Tapered Gridded Estimator (TTGE; \citetalias{Chatterjee2022}), which aims to measure the power spectrum
directly from the visibilities recorded in radio interferometric observations in the drift scan mode. The estimator is based on the TGE \citep{Choudhuri2016b} that has been widely used for analysing GMRT/uGMRT radio interferometric observations \citep{Choudhuri2020, Pal2020, Pal2022, Elahi2023, Elahi2023b, Elahi2024}.
The TGE works with gridded visibilities, a feature that makes it computationally efficient and also allows it to taper the sky response so as to suppress foreground contamination from the sources far away from the telescope's pointing direction. 
While the original TGE was developed for observations where the telescope tracks a single field. The  TTGE has been designed to work with drift scan observation. This estimator is tailored for observations with radio-interferometric arrays where the individual elements (antennas) are fixed on the ground (\eg LOFAR, MWA, HERA, CHIME, upcoming SKA-low),  and in the present work we have considered data from drift scan observations \citep{Patwa2021} with the compact configuration of MWA-II  \citep{Tingay2013}. In drift scan observations, the sky moves across the telescope's field of view with the passage of time. The TTGE allows us to follow a fixed position on the sky, namely the tracking center (TC), as it moves across the telescope’s field of view. It further allows us to taper the sky response to a small angular region around the TC so as to suppress the foreground contamination from bright sources located at large angular separations from the TC.

In \citetalias{Chatterjee2022}, we have presented the mathematical framework for the TTGE and validated this for a single frequency channel. The analysis there considers an input model angular power spectrum $C_{\ell}^M$, for which the simulated sky signal and resulting drift scan visibilities are processed through the TTGE pipeline. The estimated $C_{\ell}$  is found to be in good agreement with  $C_{\ell}^M$,  thereby validating the TTGE.  The present work validates the TTGE considering observations that cover multiple frequency channels spanning a finite bandwidth.  Here, we have considered an input model power spectrum  $P^m(k)=(1 {\rm Mpc}^{-1}/k) \, {\rm K^2}\, {\rm Mpc^3}$, for which simulated all-sky maps were used to calculate drift scan visibilities. These were processed through the TTGE pipeline to estimate the MAPS $\cl(\dnu)$ which was found to be in good agreement with the predicted $\cl^M(\dnu)$ corresponding to $P^{\rm m}(k)$ (Figure~\ref{fig:valid_cl}).  We have used a  Maximum Likelihood estimator  \citep{Pal2022, Elahi2023} to directly compute both $P(\kpp, \kpar)$ and $P(k)$ from the estimated $\cl(\dnu)$. 
The estimated $P(k)$ is found to be in good agreement with the input model $P^{\rm m}(k)$ (Figure~\ref{fig:valid_1pk}), thereby validating the TTGE for 3D multi-frequency data. 

The MWA has a periodic pattern of flagged channels, the period being $\dnu_{\rm per}  = 1.28  \,{\rm MHz}$  (Figure~\ref{fig:bl-nu}). In addition, there is the possibility of other frequency channels also being flagged in order to avoid man-made radio frequency interference (RFI), etc. 
Several earlier works which have first transformed from frequency to delay space, and then estimated the PS (e.g. \citealt{Paul2016, Li2019, Trott2020, Patwa2021})  have reported a very prominent pattern of horizontal streaks in the estimated $P(\kpp, \kpar)$ due to the periodic pattern of flagged channels present in the  MWA data.  In our approach \citep{Bharadwaj2018}, we first estimate $\cl(\dnu)$, which does not have any missing $\dnu$ even when the visibility data contains flagged channels. We then Fourier transform  $\cl(\dnu)$ to obtain a clean estimate of  $P(\kpp, \kpar)$. Considering the simulated data, we see that the missing frequency channels do not introduce any artefacts in the estimated $P(\kpp, \kpar)$ (Figure~\ref{fig:valid_3pk}).  We find that the estimated $P(\kpp, \kpar)$ are visually indistinguishable irrespective of whether there are flagged channels or not. We conclude that our estimator is impervious to the periodic pattern of flagged channels, and these do not introduce any artefacts in the estimated $P(\kpp, \kpar)$.

We have applied the TTGE to estimate the PS for a single pointing of the actual MWA data, which effectively corresponds to using the TGE.  The pointing center PC=34 corresponds to $({\rm RA,DEC)} = (6.1^{\circ},-26.7^{\circ})$   (Figure \ref{fig:sky_covered})  with an observing time of approximately $17 \, {\rm min}$. For the tapering, we have used a Gaussian window function with $\thf = 15^{\circ}$. The estimated MAPS $\cl (\Delta \nu)$,  arising mainly from foregrounds, is shown in Figure~\ref{fig:cl_sys}.  This does not appear to exhibit any prominent features which can be associated with the periodic pattern of flagged channels. However, on fitting and subtracting out a smooth polynomial in $\Delta \nu$, for several $\ell$ values the residual $\cl(\Delta \nu)$ exhibit oscillations of period $\Delta \nu_{\rm per}=1.28 \, {\rm MHz}$.  The amplitude of these oscillating residuals is approximately two orders of magnitude smaller than the total $\cl(\Delta \nu)$. 
The left panel of Figure~\ref{fig:pc4} shows the cylindrical power spectrum $P(\kpp, \kpar)$ estimated from the measured  $\cl(\Delta \nu)$. We find that the foregrounds are mainly localized within the expected foreground wedge boundary, with some leakage beyond. We notice a few horizontal streaks in the power spectrum that extend across several  $\kpp$ bins at some fixed $\kpar$ values. 

To understand the origin of these streaks, we consider Figure~\ref{fig:per_sys} where we find the power spectrum at $k_\perp=0.01\,{\rm Mpc}^{-1}$ shows spikes at a regular interval of $\Delta \kpar \sim 0.29\,{\rm Mpc}^{-1}$ which matches $\Delta \nu_{\rm per}=1.28 \, {\rm MHz}$ in $\cl(\Delta \nu)$. We attribute the spike in $P(\kpp, \kpar)$ to the periodic oscillation in $\cl(\Delta \nu)$. We also find that $P(\kpp, \kpar)$ does not exhibit such prominent spikes at the larger $\kpp$ bins. Further, such spikes are also not present in the $P(\kpp, \kpar)$ estimated from simulations, which incorporate the periodic pattern of flagged channels. 
Note, however, that the dynamic range of the present simulations is an order of magnitude lower than the actual data. In future work, we plan to carry out foreground simulations where we expect to achieve a higher dynamic range.

The origin of the small periodic oscillation in $\cl(\dnu)$, which is responsible for the spikes in $P(\kpp, \kpar)$, is not known at present. The fact that these features are not present in the simulations suggests that our estimator is impervious to the periodic pattern of flagged channels, and the spikes are not just a straightforward consequence of the missing channels. Our results seem to indicate that the spikes are possibly caused by some systematic effect in the data, which has the same period as the missing channels, periodic systematic errors in the calibration being one such possibility. We note that the streaks found here are an order of magnitude smaller than those found in earlier analyses of MWA data (e.g. \citealt{Patwa2021}). It may be possible to model and remove these artefacts by subtracting out small period components from the measured $\cl(\dnu)$ using Gaussian process regression or some similar technique \citep{Mertens2018, Elahi2023b}. We plan to address this in future work. 

We identify a rectangular region $0.05 \leq \kpp \leq 0.16 \, {\rm Mpc^{-1}}$ and $0.9 \leq \kpar \leq 4.6 \, {\rm Mpc^{-1}}$ to be relatively free of foreground contamination, spikes, and other artefacts. We use the $P(\kpp, \kpar)$ estimates in this region to constrain the EoR 21-cm signal. The $P(\kpp, \kpar)$ values, suitably scaled with the uncertainties expected from the system noise, are found to be nearly symmetrically distributed around 0. The distribution is well-fitted with a Lorentzian profile. However, the standard deviation is roughly 100 times of that expected from the contribution from system noise alone. The cause of this excess variance is not known. We have used our measurements to place a $2\sigma$ upper limit $\Delta^2(k) < (1.85\times10^4)^2\, {\rm mK^2}$ on the mean squared 21-cm brightness temperature fluctuations at $k=1 \,{\rm Mpc}^{-1}$. The present results, which are restricted to a single pointing, are limited by foregrounds and the systematics whose origins are not known to us at present. It is necessary to address these issues before combining multiple pointings. We plan to address these issues in future work. 

\appendix

\section{Preliminary results: MAPS} 
\label{app:cl}

\begin{figure*}
    \centering
    \includegraphics[width=0.95\textwidth]{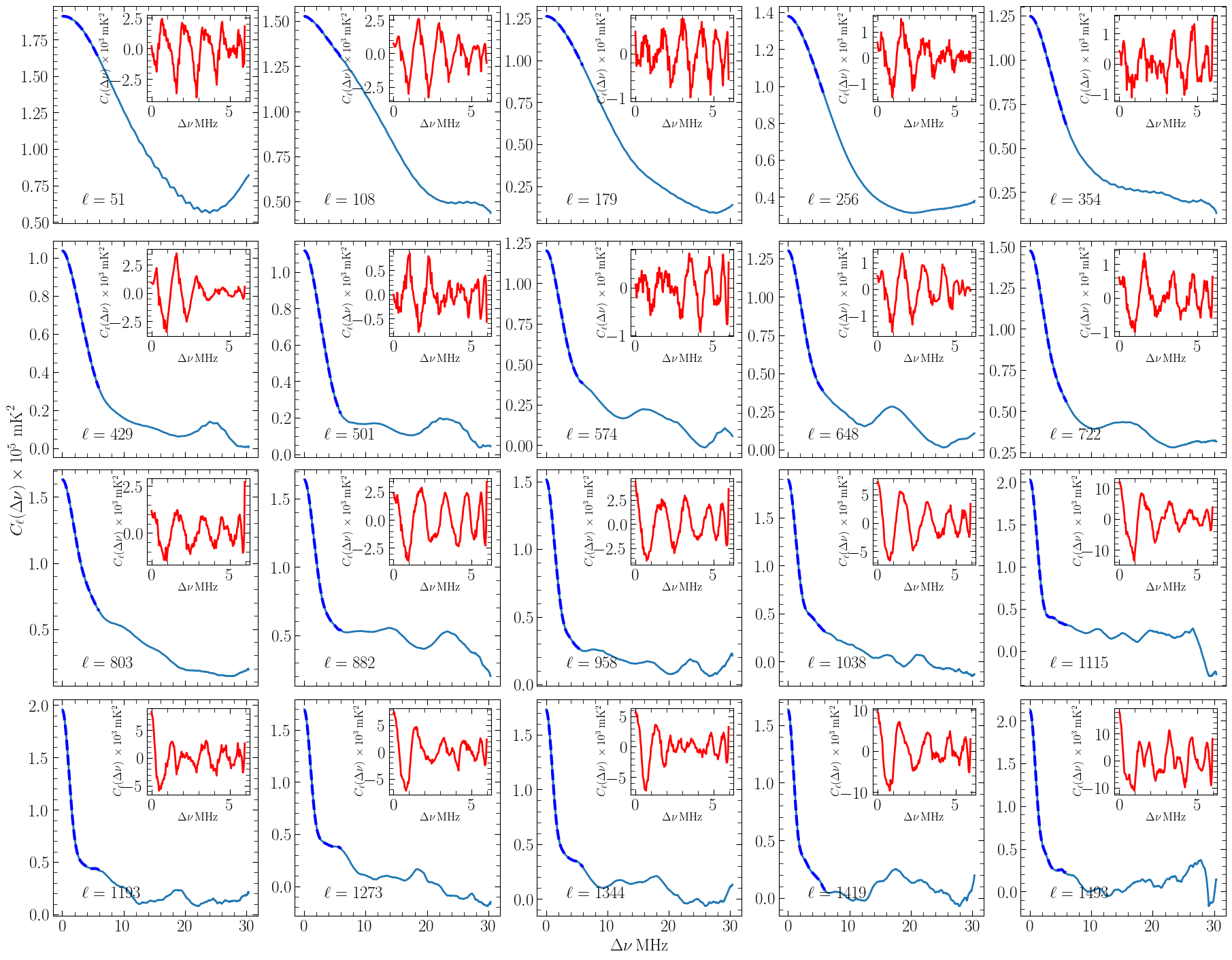}
    \caption{This figure shows $\cl(\dnu)$ for the annotated $\ell$ values. The blue dashed curves show a polynomial fit on the range $\dnu<6$ MHz. The polynomial fit is subtracted from the measured $\cl(\dnu)$, and the residual $\cl(\dnu)$ are shown in the insets (red).}
    \label{fig:cl_sys}
\end{figure*}

We have applied the TTGE on MWA drift scan observation to estimate the MAPS $\cl(\dnu)$ (Section~\ref{sec:results_data}). Figure \ref{fig:cl_sys} shows $\cl(\dnu)$ for different $\ell$ whose values are annotated in each panel. The estimated  $\cl(\dnu)$ have values $\sim 2\times10^5 \,{\rm mK}^2$ at $\dnu=0$ and it gradually decreases with increasing $\dnu$. This gradual decorrelation is a typical behaviour of foregrounds, which has a smooth frequency dependence. Further, the decorrelation becomes faster with increasing $\ell$ due to baseline migration \citep{Pal2022}. We note that point sources, which are in the sidelobes of the telescopes, introduce oscillatory features in $\cl(\dnu)$, which are not prominent in these plots. This is because the contribution from the sources within the field of view of the primary beam is much larger than those appearing through the sidelobes. We plan to identify and remove the point sources that are in the field of view of the primary beam in future work. 

We see no prominent features in the estimated  $\cl(\dnu)$ but find features in the power spectrum (Figure~\ref{fig:pc4}), which leads us to investigate the measured $\cl(\dnu)$ closely. The idea is to fit the measured $\cl(\dnu)$ with a polynomial and subtract the polynomial fit to look for any features in the residual $\cl(\dnu)$. Here we have chosen the range $\dnu<6$ MHz and fitted the estimated  $\cl(\dnu)$ with an even polynomial having 10 terms. The blue dashed curves on top of the measured $\cl(\dnu)$ show the polynomial fits, and the residual $\cl(\dnu)$ are shown in the insets (red). 

\section*{Acknowledgements}
S. Chatterjee acknowledges support from the South African National Research Foundation (Grant No. 84156) and the Department of Atomic Energy, Government of India, under project no. 12-R\&D-TFR-5.02-0700. S. Chatterjee would also like to thank Dr. Devojyoti Kansabanik, for helpful discussions.

\section*{Data Availability}
The data sets were derived from sources in the public domain
(the MWA Data Archive: project ID G0031) at \url{https://asvo.mwatelescope.org/}. 

\bibliography{mylist.bib}
\label{lastpage}
\end{document}